# Low-voltage coherent electron imaging based on a single-atom electron source

Wei-Tse Chang[1], Chun-Yueh Lin[1,2], Wei-Hao Hsu[1,3], Mu-Tung Chang[1], Yi-Sheng Chen[1], En-Te Hwu[1], and Ing-Shouh Hwang[1,3]

[1]Institute of Physics, Academia Sinica, Nankang, Taipei, Taiwan

[2]Department of Physics, National Taiwan University, Taipei, Taiwan

[3]Department of Materials Science and Engineering, National Tsing Hua University, Hsinchu, Taiwan

E-mail:ishwang@phys.sinica.edu.tw,

**Abstract**

It has been a general trend to develop low-voltage electron microscopes due to their high imaging contrast of the sample and low radiation damage. Atom-resolved transmission electron microscopes with voltages as low as 15-40 kV have been demonstrated. However, achieving atomic resolution at voltages lower than 10 kV is extremely difficult. An alternative approach is coherent imaging or phase retrieval imaging, which requires a sufficiently coherent source and an adequately small detection area on the sample as well as the detection of high-angle diffracted patterns with a sufficient resolution. In this work, we propose several transmission-type schemes to achieve coherent imaging of thin materials (less than 5 nm thick) with atomic resolution at voltages lower than 10 kV. Experimental schemes of both lens-less and lens-containing designs are presented and the advantages and challenges of these schemes are discussed. Preliminary results based on a highly coherent single-atom electron source are presented. The image plate is designed to be retractable to record the transmission patterns at different positions along the beam propagation direction. In addition, reflection-type coherent electron imaging schemes are also proposed as novel methods for characterizing surface atomic and electronic structures of materials.

# 1. INTRODUCTION

Imaging and characterizing two-dimensional (2D) materials, such as graphene and metal dichalcogenides, and low-atomic-number materials, such as organic or biological molecules, are of great importance in science and technology. The low contrast and possible radiation damage of these materials pose challenges to conventional electron microscopy, which is mainly based on electron beams of 100 keV or higher. In recent years, several research groups have developed low-energy transmission electron microscopes (TEMs) to detect the atomic and electronic structures of thin samples (Sasaki et al. 2010; Krivanek et al. 2010; Kaiser et al. 2011; Bell & Erman 2013]. This is because the scattering cross section of atoms increases as electron energy decreases. Another aim is to reduce radiation damage. Materials can be damaged via the knock-on mechanism at voltages higher than a threshold (Scherzer, 1970; Glaeser 1978), which is approximately 60 kV for carbon atoms (Rose, 2009). TEMs with voltages as low as 15-40 kV (Kaiser et al. 2010; Sasaki et al. 2100; Dellby et al. 2011; Sasaki et al. 2014) have been demonstrated to achieve atomic resolution. However, a beam-induced chemical etching effect was observed on graphene samples at electron energy ranging from 20 to 80 keV in an ultra-high vacuum (UHV) environment (Meyer et al. 2012). In addition, the bonding energy of atoms at surfaces is considerably lower than that of atoms in the bulk. Thus, surface or edge atoms may be displaced or etched at low voltages (Suenaga et al. 2011). These effects underline the importance of developing electron microscopes that can be operated at lower voltages for imaging low-dimensional structures and low-atomic-number materials. Continuous advancement in aberration correction techniques has been the key factor in achieving high-resolution imaging at even lower electron energies. However, such techniques become extremely challenging and complicated as the operating voltage reaches levels lower than 10 kV. A 5-kV electron microscope developed by Delong Instruments (LVEM5) achieved a resolution of approximately 2 nm (Drummy et al. 2004), which is far from atomic resolution. Nevertheless, the microscope provides high-contrast imaging of organic materials.

An alternative approach is phase retrieval imaging or coherent imaging (Fienup, 1982; Miao et al. 1998; Miao et al. 2000; Humphry et al. 2012; Gabor, 1948; Miao et al. 1999). Figure 1 illustrates a schematic of phase retrieval imaging. Here we assume that an isolated thin object with a width of $w$ is illuminated with a coherent beam wider than the object or that an extended object is illuminated with a coherent beam with a width of $w$. Conventional transmission-type microscopes typically contain a post-object lens system, which also introduces a certain degree of spherical aberration, used to form magnified images of an object. A post-object lens is not required in coherent imaging and a 2D detector, or an image plate, is placed at a certain distance behind the object to record the transmission patterns. The wave scattered from an object contains information about the object's structure. However, detectors are sensitive to only the intensity of the wave, thus leading to the loss of the phase information. This is known as the "phase problem". Several phase-retrieval methods have been developed for retrieving the lost phase. By combining the phase and intensity information, the full information of a structure can be reconstructed through an algorithm. Therefore, the algorithm can be viewed as a virtual lens to form images of an object. The algorithm is free from aberrations typically observed in real lens systems. Moreover, both the modulus and phase images can be obtained. In particular, the phase images contain quantitative information of electric and magnetic fields in a material (Liche & Lehmann 2008).

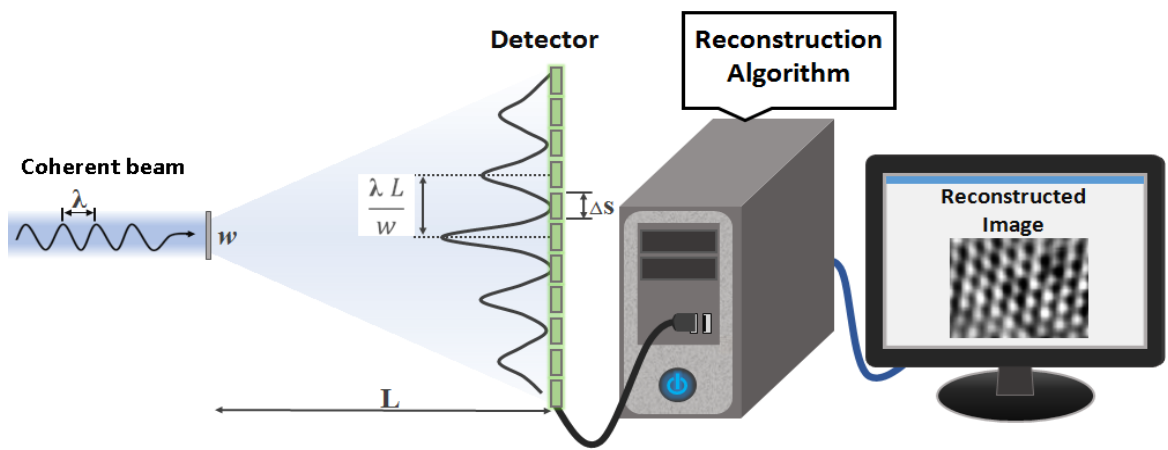

Figure 1 Schematic of phase retrieval imaging. A coherent source illuminates a thin object of lateral size $w$, and the transmission pattern is recorded by a detector at a distance $L$ behind the sample. The transmission patterns can be used for calculations in a reconstruction algorithm to obtain the images of the object.

In phase retrieval imaging techniques, there is an oversampling requirement for successful reconstructing the object image from a transmission pattern, which can usually be considered a diffraction pattern. As illustrated in Figure1, the finest Young's fringes in the diffraction pattern have a period of $\lambda L/w$ (Spence et al. 2004), where $\lambda$ is the wavelength of the coherent beam and $L$ is the distance between the object and an electron detector screen. The sampling theorem requires at least two samples per period of the intensity; therefore, the pixel spacing of the detector, $\Delta s$, should be smaller than $\lambda L/2w$, the inverse of the Nyquist frequency. This places a limitation on the object size or the illumination width, $w < \lambda L/2\Delta s$. Theoretical calculations are essential for image reconstruction in phase retrieval techniques. Several review articles (Fienup, 1982; Latychevskaia et al. 2012; Rodenburg, 2008; Miao et al. 2012) have addressed reconstruction methods and such theoretical methods are beyond the scope of the current work. In this work, we focused on the experimental realization of phase retrieval imaging by using low-energy electrons. The ultimate goal is to achieve high-contrast and high-spatial-resolution imaging of thin materials or molecules under low-dose conditions.

According to the concept of phase retrieval imaging, several microscopic techniques, such as holography (Gabor, 1948), coherent diffractive imaging (CDI) (Miao et al. 1998; Miao et al. 2000; Miao et al. 1999; Liche & Lehmann 2008; Spence et al. 2004; Latychevskaia et al.2012; Rodenburg, 2008; Miao et al. 2012; Zuo et al. 2003; Weierstall et al. 2002; Morishita et al. 2008; Huang et al. 2009; Caro et al. 2010; Kamimura et al. 2011; Kamimura et al. 2008; Kamimura et al. 2010; Marathe et al. 2010), ptychographic CDI (Humphry et al. 2012; Faulkner & Rodenburg 2004 a; Faulkner & Rodenburg 2004 b; Faulkner & Rodenburg 2005), and keyhole CDI (Fresnel CDI) (Abbey et al. 2008; Williams et al. 2006), have been successfully implemented using x-ray or electron waves. Weierstall *et al.* (Weierstall et al. 2002) first demonstrated coherent electron diffractive imaging. A double-hole image was reconstructed with a resolution of 5 nm according to diffraction patterns acquired at 40 kV. Several electron beam CDI experiments were subsequently implemented and atomic level structures such as carbon nanotubes and Au nanocrystals were successfully reconstructed (Miao et al. 1998; Zuo et al. 2003; Morishita et al. 2008; Huang et al. 2009; Caro et al. 2010; Kamimura et al. 2011). Most of the experiments were conducted using commercial TEMs with a voltage of 100 kV or higher. Kamimura et al. executed 10-kV electron diffractive imaging by modifying a conventional scanning electron microscope (SEM) and successfully resolved the wall spacing of 0.34 nm of a multiwalled carbon nanotube (Kamimura et al. 2010). They later demonstrated diffractive imaging of a single-walled carbon nanotube (SWCNT) with a resolution of 0.12 nm by using 30-kV electron beams (Kamimura et al. 2011). In this single-exposure CDI technique, individual nano-objects were imaged. The object size is limited by the oversampling condition. Rodenburg et al. proposed ptychographic CDI for imaging extended objects (Humphry et al. 2012; Faulkner & Rodenburg 2004 a; Faulkner

& Rodenburg 2004 b; Faulkner & Rodenburg 2005). This scheme was demonstrated recently by using a SEM optical column to generate a convergent electron beam and to scan the beam on the sample (Humphry et al. 2012). Diffraction patterns of a number of overlapped areas were recorded by performing a raster scanning of the electron beam. Images of the sample of an extended region were reconstructed and 0.236 nm atomic plane fringes of gold nanoparticles were resolved at a voltage of 30 kV. An advantage of ptychographic CDI is that the entire imaged area is no longer limited by the oversampling condition. The oversampling condition must still be met by focusing the beam to a sufficiently small size on the sample for recording each diffraction pattern.

Although current electron-beam based coherent imaging techniques have achieved atomic level structures, such techniques were applied to nano-materials with good structural order, such as carbon nanotubes or gold nanocrystals. A more challenging and vital goal is to resolve the atomic structures of non-periodic structures such as individual proteins and other biological molecules. However, this objective has not yet been achieved. In principle, coherent electron imaging has a potential to achieve this goal. The current study proposes new schemes to lead coherent imaging closer to achieving this goal. Two directions were adopted: operation of electron microscopes at voltages lower than 10 kV and the use of highly coherent single-atom electron sources.

In coherent imaging, high spatial frequency components of the elastically scattered object wave travel at high angles. The resolution is determined according to the angle that the detector subtends at the specimen. The high-angle diffracted patterns must be recorded to reconstruct correctly the high-frequency spatial components of the object image. A major advantage of low-energy electrons over high-energy electrons and x-ray is that the cross-sections of the interaction with the atomic potentials are rather large, thus enabling the diffraction patterns to have favorable signal-to-noise ratios even at high scattering angles, and the high-resolution images can be reconstructed.

We adopted a single-atom electron source because the spatial coherence (or lateral coherence width) of an electron source increases as the source size decreases. We used noble-metal covered W(111) single-atom tips (SATs) because they can be reliably prepared and regenerated in vacuum (Fu et al. 2001; Kuo et al. 2004; Chang et al. 2009). Their pyramidal structures are thermally and chemically stable, ensuring their long operation lifetime. Previous studies have reported that the electron beams emitted from such SATs demonstrate high brightness and full spatial coherence (Fu et al. 2001; Kuo et al. 2004; Chang et al. 2009), which are ideal for phase retrieval imaging.

To achieve local coherent illumination to meet the oversampling condition, two schemes can be used. The first one is based on a lens-less imaging technique. A small illumination area can be achieved if the specimen is placed sufficiently close to a sharp emitter. The second one entails using a lens system to focus the electron beam to a sufficiently small spot on the sample. Each of these schemes has its advantages and disadvantages, and these are discussed subsequently in this paper. In addition to Section 1, which presents the introduction, Section 2 describes the sample and emitter preparation process. Section 3 presents the lens-less scheme for transmission-type coherent imaging. This section also presents preliminary results on monolayer graphene. Section 4 proposes a low-kV transmission-type scheme comprising several imaging modes. The image plate is designed to be retractable to record the transmission patterns at different positions along the beam propagation direction. Section 5 presents the features of single-atom electron sources and the strategy for building electron microscopes according to such electron sources. Section 6 proposes reflection-type schemes for coherent electron imaging; these schemes are used for characterizing atomic and electronic structures on surfaces. Section 7 presents the conclusion.

**2. Samples and emitter preparation**

2.1 Sample preparation for transmission-type coherent electron imaging

Because electrons with energies less than 10 keV have low penetration depths, only materials with a thickness equal to or less than 3-5 nm can be studied. The penetration depth should be sufficient for studying few-layer materials and many nanostructures and organic molecules. A membrane (such as Si, $Si_3N_4$) fabricated using microelectromechanical system technology (figure 2a) or a holey carbon film typically used as a TEM sample holder can serve as a sample holder. The sample holder should be electrically conductive, which can be achieved either through doping or coating with a thin metal film. If the electron beam illuminating on the sample is too wide, a membrane that has a hole size fulfilling the oversampling condition can be used. To ensure that only one holed region is illuminated, the spacing between neighboring holes should be wider than the beam size. One-dimensional (1D) nanostructures (such as carbon nanotubes and nanowires) or organic molecules (such as polymer and DNA molecules) can be placed across the holes of the support membrane (Latychevskais & Fink 2009) and the freestanding part can then be studied (Figure 2b). Figure 2c illustrates a 2D material covering the holed regions. Methods should be developed to transfer 2D materials to the sample holder. Organic molecules and nanostructures can be deposited on free-standing 2D material, such as a clean suspended graphene, for coherent electron imaging (Figure 2d). The structures can also be sandwiched between two few-layer materials such as graphene (Zaniewski et al. 2013). A stiff 1D structure, such as a carbon nanotube or a nanorod, can be attached to the apex of a sharp metallic needle such that a portion of the structure sticks out of the tip and can be studied (Figure 2e). Nanostructures can also be placed at the apex (Figure 2f).

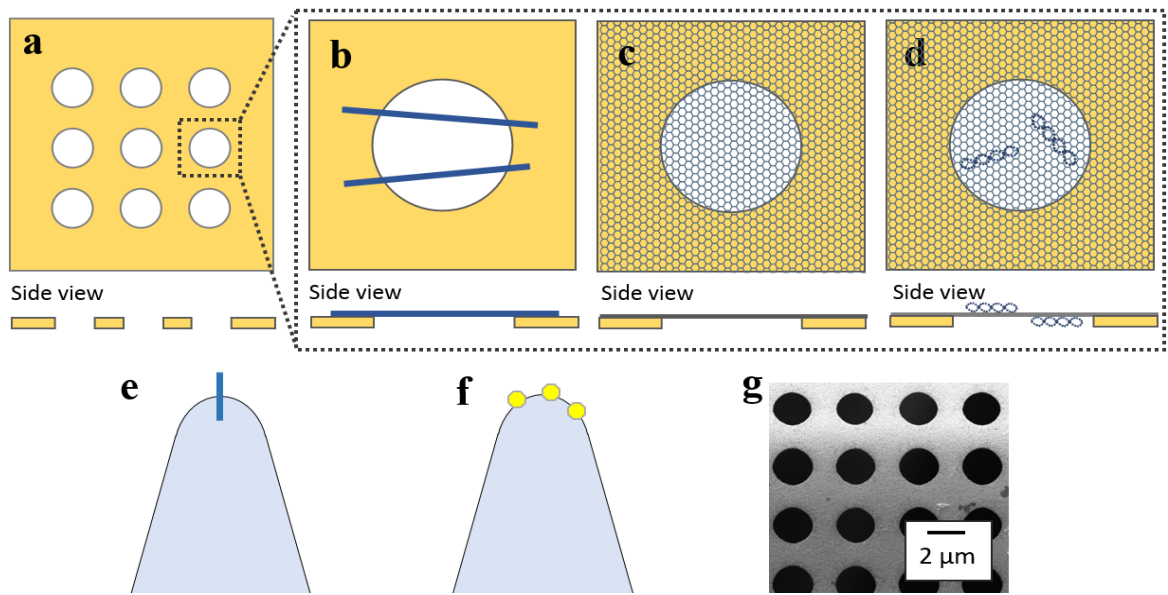

Figure 2  Sample preparation for low-energy electron coherent imaging. (a) Holey membrane. The bottom inset shows the side view of this membrane. (b) 1D materials across a holed region. The bottom inset shows the side view of a holed region. (c) 2D materials covering a holed region. The bottom inset shows the side view of a holed region. (d) Molecules or nanostructures on the suspended 2D materials. The bottom inset shows the side view of a holed region and the molecules can be on either side of the suspended 2D material. (e) 1D structure sticking out from a tip apex. (f) Nanostructures adsorbed at the tip apex. (g) Scanning electron micrograph of a part of the microfabricated $Si_3N_4$ membrane. The width of each hole is 2 μm.

In this work, a graphene sheet supported on a gold-coated $Si_3N_4$ membrane containing periodic holes was prepared for conducting experiments (see Section 3). Figure 2g depicts a scanning electron micrograph of a part of the membrane. A CVD graphene sheet on a polycrystalline copper foil (ASC Material) was transferred onto the membrane by using a polymer-free transferring method (Lin et al. 2014). The copper foil was subjected to chemical wet-etching on a clean Petri dish by using 0.5 M Iron(III) chloride solution ($FeCl_3$). After complete removal of the copper foil, a monolayer graphene sheet floated on top of the etchant. The etchant was then gradually replaced with deionized (DI) water at a rate of 1 mL/min. After the etchant was completely replaced, a $Si_3N_4$ membrane was placed below the floating graphene sheet. The DI water was gradually drained out to lower the graphene sheet until it rested on the membrane. After the transferring process, the sample was heated on a hot plate from 30°C to 110°C at a rate of 5°C/min to remove water on the sample.

2.2 Preparation of single-atom emitters

To prepare a noble-metal covered W(111) SAT (Kuo et al. 2004), a W(111) wire was first etched into a tip with a radius of 20-40 nm (Chang et al. 2012). After the tip was electroplated with a thin noble-metal film (Kuo et al. 2004), the emitter was placed in a UHV chamber and then annealed until a pyramidal SAT formed. The tip formation process can be monitored by observing the electron field emission pattern. Previous studies have adequately characterized the relationship between the FEM pattern and the corresponding FIM images of a SAT and the field emission current stability (Hwang et al. 2010; Kuo et al. 2006). In the case of Ir covered W(111) SAT, the tip annealing temperature is approximately 1100 K.

3. Lens-less transmission-type coherent imaging

In 1939, Morton and Ramberg first demonstrated lens-less electron microscopy by using a type of electron projection microscope (PPM) (Morton & Ramberg 1939), a shadow microscope, in which an object is placed between a field emission electron point source and a screen. The magnification of the shadow image is equal to the ratio of the source-screen distance to the source-object distance. In 1990, Stocker, Fink and co-workers improved this type of PPM by using a highly coherent electron point source and successfully imaged carbon fibers of 10-20 nm (Fink et al. 1990a; Fink et al. 1990b). Several researchers have built similar setups and demonstrated that this type of PPM could easily yield magnifications up to a factor of $10^6$ without using lenses (Binh et al.

1994; Park et al. 1999; Binh et al. 2000; Spence et al. 1994; Fink et al. 1999; Eisele et al. 2008; Weierstall et al. 1999; Longchamp et al. 2012; Mutus et al. 2011). Typical electron energies used in PPMs range from 30 to 500 eV, considerably lower than those used in modern high-voltage electron microscopes. Images obtained with this type of microscopes are free from spherical aberrations. However, the optimal spatial resolution obtained has never reached a level better than 2 nm (Beyer & Gölzhäuser 2010). In a previous study, a single-atom-electron-source based PPM was used to image a suspended and isolated SWCNT at various tip-sample separations (Hwang et al. 2013). Through numerical simulations to fit the recorded interference patterns, it was concluded that the interference patterns at small tip-sample separations were dominated by the biprism effect because of a considerable charge density induced on the nanotube. This explains why images obtained using PPMs thus far have never achieved a resolution close to atomic level.

3.1 New scheme for lens-less coherent imaging by using low-energy electrons

Previous PPMs were not designed to record high-angle transmission patterns because a large sample-detector separation can yield images of high magnification. Thus the angle of the detector subtended at the sample is narrow; this limits the resolution of the reconstructed images. For example, a typical sample-detector separation is approximately 10 cm and the detector is typically a multichannel plate (MCP) with a diameter of nearly 8 cm, yielding an angle of approximately ±22°. If the detector screen is placed closer to the sample, the magnification of the PPM images is reduced and the pixel spacing in the detector screen may not fulfill the oversampling requirement. Figure 4a shows a new scheme with a MCP-screen assembly that can be moved along a rail to record the transmission patterns at various positions along the beam propagation direction. This retractable detector can be moved to the farthest position behind the sample to record high-magnification images and to a point close to the sample to record wide-angle transmission patterns. This additional degree of freedom can enable the development of theoretical methods for reconstructing high-resolution images.

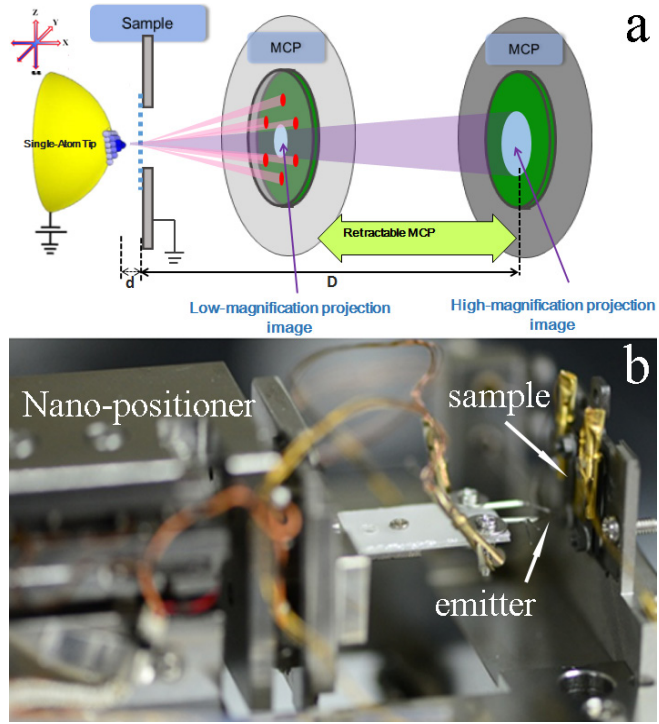

Figure 4 New schematic for lens-less coherent imaging involving low-energy electrons. This system is a low-energy electron PPM with a retractable image detector. The electron beam is extracted from the topmost atom of an SAT. The magnification of the projected image on the MCP is (D + d)/d, where D is the sample-detector separation and d is the tip-sample separation. (b) Photo of a PPM-based electron microscope, which is firmly attached to a 8" flange. The MCP-screen assembly (not shown here) mounted on a rail is attached to another 8" flange on the opposite side of the vacuum chamber.

Figure 4b illustrates a photo of our proposed PPM-based electron microscope. The microscope is housed in a UHV chamber with a base pressure of approximately $1\times10^{-10}$ Torr. A SAT is mounted on a three-axis piezo-driven positioner (Unisoku, Japan) with a 5 mm traveling distance in each direction. The tip can be moved closer to the sample to get a high-resolution PPM image. An MCP-screen assembly (diameter = 77 mm, F2226-24PGFX from Hamamatsu) is mounted on a rail such that it can be moved to various positions located 3-14 cm behind the sample. A camera (Andor Neo 5.5 sCMOS, 16-bit, 2500 pixel× 2000 pixel) adapted with a camera head (Nikon AF Micro-Nikkor 60 mm f/2.8 D) is placed behind the screen outside the UHV chamber to record the images on the screen.

Figure 5(a) shows a PPM image of a monolayer CVD graphene, whose preparation procedure is described in Sec. 2.1, acquired at the tip bias of 480 V when the MCP-screen assembly was positioned at 13 cm behind the sample. Several dark stripes can be observed in this image, and these are attributed to the overlapping of graphene layers belonging to different domains. The central parts of the images are brighter than the surrounding areas due to the narrow Gaussian distribution of the beam intensity (FWHM divergence between 3° and 6°) for the single-atom electron sources (Chang et al. 2009). As the emitter approached the sample, the area of the sample [Figure 5(a), blue circle] was further magnified [Figure 5(b)]. The tip further approached the sample to magnify the area outlined by a blue circle in Figure 5(b). Subsequently, the MCP-screen assembly was moved to a position 3.5 cm behind the sample to record the transmission patterns. Figure 5(c) illustrates one of the recorded images, indicating that the central pattern, which is a projection image or a hologram of the sample, exhibits a considerably smaller size on the screen; in addition, six darker patterns surrounding the central pattern are indicated in this figure. They are diffraction patterns related to a single domain of the monolayer graphene. Figure 5(d) illustrates each of these patterns, which are numbered and shown individually with enhanced contrast. Good correlation between the central patterns and the surrounding diffraction patterns can be identified, indicating that the diffracted beams contain real-space information. The central and diffracted patterns resemble the bright-field images and multiple dark-field images, respectively, in selected-area diffraction (SAD) of TEM when the electron beam is defocused on the sample. An example of SAD of TEM can be seen in Figure 11.14 of Ref. (Zangwill et al. 1998). The differences are that no lens is used here and the electron energy is lower than that in TEM by approximately three orders of magnitude.

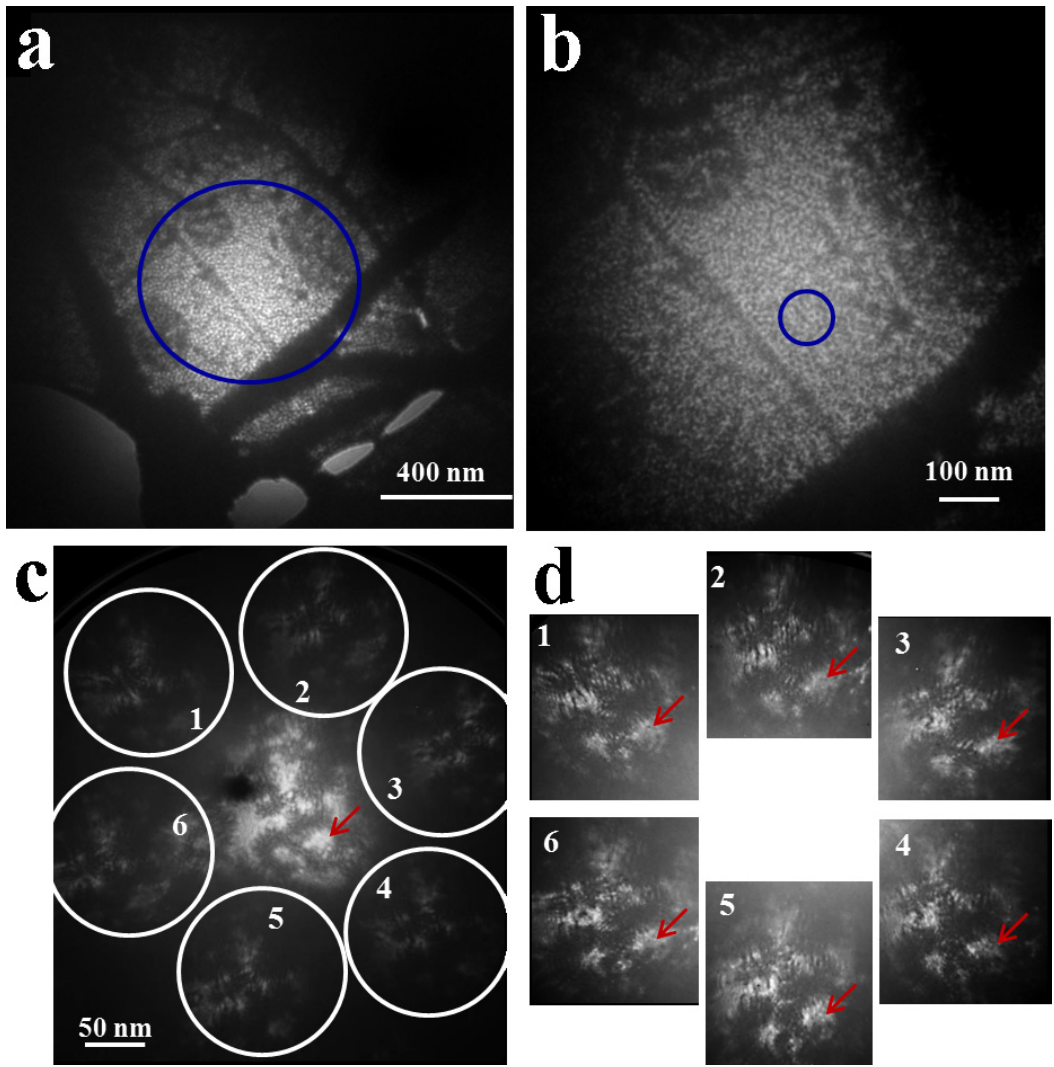

Figure 5  Transmission patterns of a monolayer CVD graphene recorded with a PPM-based electron microscope. (a) PPM image of a freestanding monolayer graphene recorded at D=13 cm. The electron energy is 480 eV. (b) PPM image of the area marked with a blue circle in (a) recorded at D=13 cm. The electron energy is 430 eV. (c) Diffraction patterns of the graphene sample recorded at D=3 cm when SAT approaches the blue circle area marked in (b). The electron energy is 270 eV. (d) Six dark-field patterns with enhanced contrast are shown. Red arrows indicate a specific feature for comparing the central bright-field image with the surrounding dark-field images. The magnification of the dark-field images is the same as that of the central bright-field image.

The bright-field and dark-field images illustrated in Figure 5 shows numerous small features with different degrees of transparency, indicating the presence of adsorbates on the monolayer graphene. The mean free path of electrons in the energy range of 20-500 eV is less than 1 nm (Germann et al. 2010). Thus, most adsorbates on the monolayer graphene are probably molecules approximately one or two atomic layers thick. Because we adopted a polymer-free transferring method, the adsorbed molecules may have originated from organic contaminations in our laboratory environment. Adsorption of residual gases on graphene in our vacuum system may also contribute to some of the adsorbates; however, it plays a minor role because of the low base pressure of the UHV chamber. Experimentally, the unknown contamination adsorbates on the graphene samples may seriously affect the process of determining the atomic structures of desired molecules. Thus, developing methods for removing contaminants from the graphene support is imperative for preparing and characterizing individual molecules on a clean graphene support. Several graphene transferring methods (Lin et al. 2014; Wang et al. 2013; Huang et al. 2014) have been reported recently. Additional treatments approaches for removing residual contaminations by annealing in gas environment (Lin et al. 2014; Huang et al. 2014), catalysis-assisted annealing (Algara-Siller & Lehtinen 2014), and dry-cleaning method (Longchamp et al. 2013a) have been introduced. Raman spectroscopy and TEM have been used to characterize the cleanness of the samples. However, only average information of selected areas on the samples can be detected from the common Raman spectroscopy. Conventional TEMs provide low-contrast images and may have chemical etching effects for adsorbates on graphene samples (Meyer et al. 2012). The current lens-less low-energy electron diffractive imaging may serve as an effective and non-destructive means of examining the adsorbates on graphene.

Figure 6 shows transmission patterns of monolayer graphene recorded at four sample-detector separations while the tip-sample separation is maintained constant. This graphene sample has been annealed at 310 °C for 12 hours under UHV and the number density of contamination is considerably reduced. The transmission patterns recorded at various distances may provide additional information for reconstructing the atomic structures of the sample.

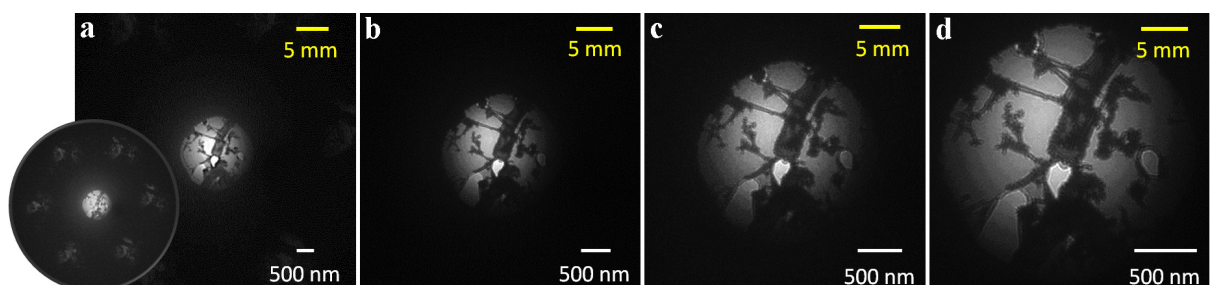

Figure 6   PPM images of the same area of a monolayer graphene recorded at the sample-detector separation of (a) 54 mm, (b) 83 mm, (c) 113 mm, (d) 143 mm. The electron energy is 360 eV.

The yellow scale bar on the upper-right hand side indicates the length scale of the screen and the white scale bar on the lower-right hand side indicates the length scale of the object. The inset in (a) is the same diffraction pattern as in (a) but plotted on a log-intensity scale to show the dark-field intensity.

3.2 Advantages and limitations of the lens-less methods

One of advantages of the lens-less scheme is that the system is very simple and the images are not affected by spherical aberration. Additional advantages include high imaging contrast and low radiation damage (Germann et al. 2010). However, this scheme has some drawbacks. First, the electron energy cannot be selected independently. Because the sample serves as an extractor for the emitter, the electron energy is limited by the tip-sample separation, which also determines the magnification of the images. Second, the small mean free path of electrons in the energy range of 20-500 eV limits the applicability of the scheme to samples less than 1 nm thick. Thus, only a few samples can be studied. For example, the diameter of a double-stranded DNA is approximately 2 nm, indicating the DNA would be opaque to electron beams in this energy range. Third, the sharp emitter may be damaged easily when the emitter is close to the specimen. Arcing is the most serious problem, and it occurs occasionally. In addition, the tip may contact the specimen momentarily during the tip approaching or tip movements to search for interesting areas. Because a voltage bias exists between the tip and the specimen, any momentary contact can lead to serious tip damage.

**4. Transmission-type coherent imaging with a focusing lens system**

Longchamp et al. recently used a micro-lens (Steinwand et al. 2010) to image a region of freestanding graphene sheet with diameter of 200 nm, and they achieved a resolution of 0.2 nm by combining coherent diffraction and holography with low-energy electrons (236 eV) (Longchamp et al. 2013b). This is the lowest electron energy reported for phase retrieval imaging. However, the penetration depth of electron beams of such low energies is less than 1 nm; thus such beams are not suitable for imaging most biological molecules. The inelastic mean-free path of electron beams exceed 1 nm as the electron energy increases to a level greater than 1 keV. Therefore, developing low-kV coherent electron imaging technique is imperative.

We propose several imaging modes for transmission-type low-kV coherent electron imaging, as schematically illustrated in Figure 7. These schematics are similar to the schematic of the PPM shown in Figure 4(a), but a lens system is added between the emitter and the sample in these schemes. A triode electron gun design is adopted in these schemes; this design enables independently controlling both emission current and electron energy (Spence et al. 2004). The first anode (extractor) is used to extract electrons from the SAT, and the second anode is used to accelerate the electron beam to 1-10 kV. An objective lens is used to focus or collimate the electron beam for different imaging modes. Figure 7(a) illustrates a CDI scheme involving a parallel beam for illuminating a sample. The challenge is to reach a sufficiently small beam diameter on the sample. For a detector with a pixel spacing of 20 μm placed at 10 cm behind the sample, the oversampling condition is that the focused beam size must not be wider than 100 nm for a 1 kV electron beam. Here, a retractable image plate, an MCP-screen assembly, is used to record the transmission patterns at various positions (Figure 4a). Both single-exposure CDI and Ptychographic CDI (Humphry et al. 2012) can be implemented. Ptychographic CDI entails scanning the electron beam with a deflective octupole or moving the sample through a piezo-scanner as in scanning tunneling microscopy (STM).

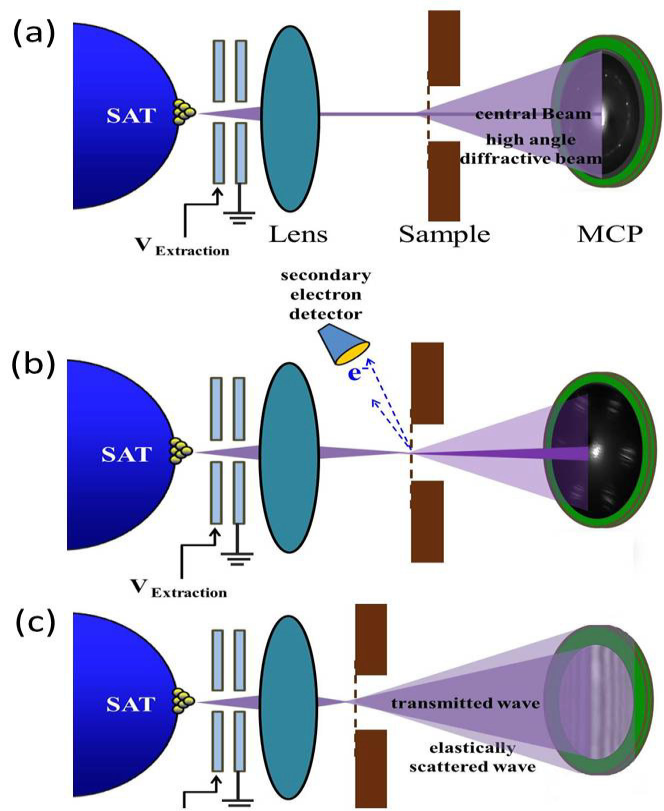

Figure 7 Three imaging modes for low-kV coherent electron imaging. The MCP-screen

assembly mounted on a rail can be moved to various positions to record the transmission patterns. (a) CDI with a parallel beam to illuminate the sample. (b) CDI with a slightly convergent beam to illuminate the sample and the transmission patterns are recorded. It also enables operation of an SEM if the secondary electrons are collected by an electron detector. (c) In-line holography. A focused electron beam with a wide converging angle is used to produce a hologram of a suspended object.

Figure 7(b) illustrates a CDI scheme that includes a slightly convergent beam for illuminating the sample. In this scheme, a spot size fulfilling the oversampling requirement can be achieved more easily than in that depicted in Figure 7(a). A single-exposure CDI or ptychographic CDI can be implemented. In addition, secondary electron microscopy images can be obtained if a channeltron or an electron detector is placed to collect the secondary electrons emitted from the sample. Figure 7(c) illustrates an in-line electron holography scheme, which is the same as that originally proposed by Gabor (Gabor, 1948). In this case, an object is illuminated with a focused electron beam of a wide converging angle in front of or behind the object. The focused point is used as the point source to produce holograms of a suspended thin object on the image plate or an MCP. A wide converging angle is required because the non-scattered reference wave must cover almost the entire detector area to extract the phase information of the object wave at high angles.

## 5. Features and strategy for single-atom electron sources

Compared with typical cold field emission electron sources, a single-atom electron source has a considerably smaller source size and narrower beam divergence. The narrow beam (2° to 3° in FWHM) renders the alignment of the electron beam a critical factor. Design for fine alignment of the source is essential for effective extraction and efficient focusing of the electron beam. Demagnification may not be required for a single-atom electron source; thus, the condenser lens can be removed, considerably shortening the lens system. In case of no demagnification, mechanical vibrations can lead to an increase in the source size, thus degrading the spatial coherence of the source. To gain the real benefits of the smallest source size, a rigid design with effective vibration isolation is required. Because the electron microscope is operated only at low kV, establishing a small and compact lens system is advantageous; this is because such a system is less susceptible to mechanical vibrations, thermal and mechanical drifts, etc. The design of UHV-STM can also be adopted; this design has been used to routinely achieve atomic resolution on flat sample surfaces. Furthermore, electron emitted from a SAT may exhibit the antibunching effect (Kiesel et al. 2002). Thus, the Boersch effect, which is caused by space charge interactions among the emitted electrons and can lead to an increase in the energy spread and the size of the electron beam as it moves down the column, should be considerably less serious than electron beams produced from normal tungsten cold field emitters. Considering the unique emission characteristics of single-atom emitters, the electron gun and lens system must be optimized.

The system can demonstrate best performance under the UHV condition because of the heat treatment for formation of a SAT and stability of field emission current. In addition, a clean sample surface can be prepared and maintained only under UHV condition because the adsorption rate of residual gases under this condition would be minimal. Piezo-actuators, which have been widely employed in UHV-STM systems, can be used to establish a rigid and compact system and to achieve precision positioning and alignment. Here, we propose a system comprising multi-axis piezo-actuators to position the SAT relative to the lens system. Figure 8(a) illustrates a schematic of the proposed system. The emitter, lens system, and sample are integrated on a stainless steel stage of 80 mm × 90 mm, which is supported by a viton-stage for vibration isolation. Figure 8(b) shows a photo of the mechanical stage (excluding the MCP). The emitter is mounted on a five-axis piezo-positioner (X-Y-Z linear displacement and 2D angular tilt), the lens system is fixed on the stage, and a sample stage is mounted on another three-axis piezo-positioner. An electrostatic lens is preferred because it is UHV compatible and relatively simple to fabricate. The entire system is placed inside a UHV chamber sitting on a vibration-isolation table.

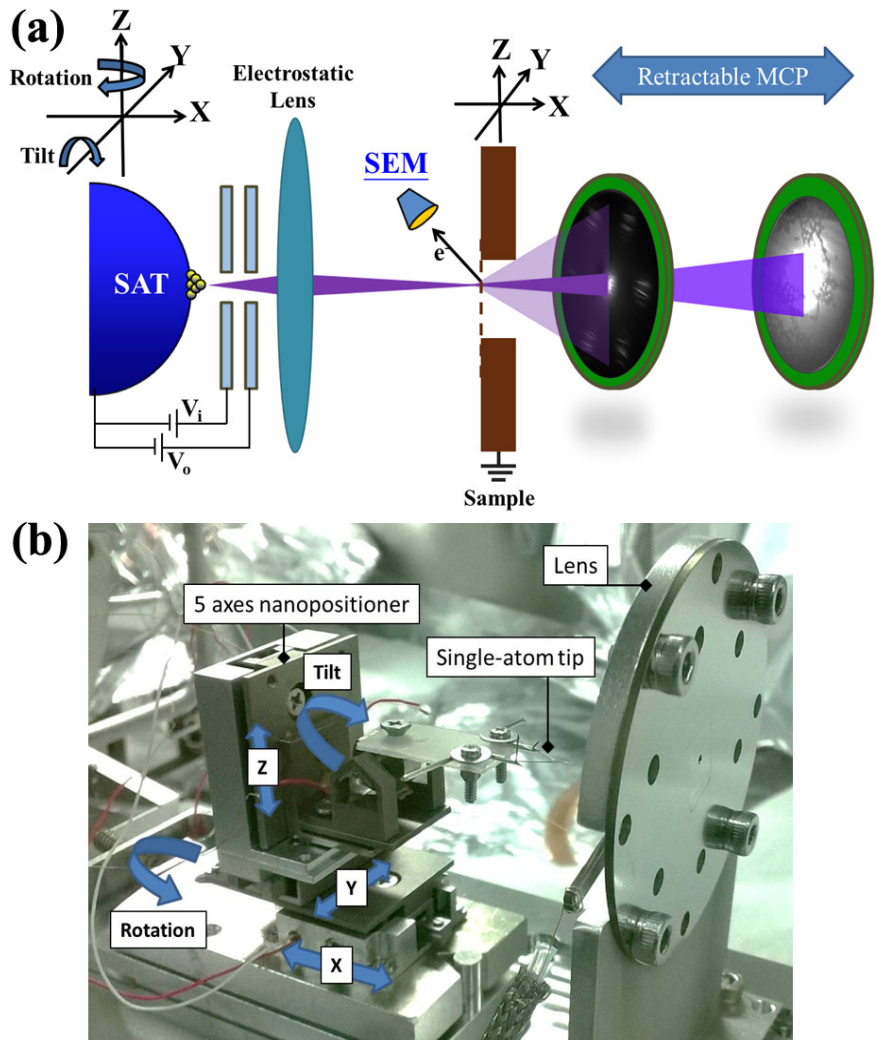

Figure 8  (a) Schematic of transmission-type low-kV coherent electron imaging. (b) Photo of a low-kV electron microscope.

The retractable image plate not only enables us to record transmission patterns at different positions, but also helps us in aligning the electron beam. Figure 9 shows the alignment of the narrow electron beam emitted from the SAT. The misalignment may occur in angular and/or lateral directions. To monitor the angular misalignment of the emitter, the MCP is displaced along the beam propagation direction. A tilted beam causes the center of the emission spot on the MCP to shift when the MCP is moved forward or backward [Figure 9(a)]. The two-axis tilting of the beam can be compensated by adjusting a 2D angular positioner [Figure 9(b)]. The lateral misalignment of the electron beam relative to the long axis of the lens system can be monitored by actuating the focusing lens. Figure 9(c) illustrates a shift of the emission spot before and after activating the lens; this shift is attributed to the lateral misalignment of the beam. This alignment can be compensated by adjusting the lateral position of the tip on the Y-Z plane by using a 2D linear positioner. Finally, an adequately aligned SAT can be obtained [Figure 9(d)]. The distance between the SAT and the extractor can be adjusted by placing an additional linear positioner in the X direction; this provides an extra degree of freedom in optimizing the focusing condition.

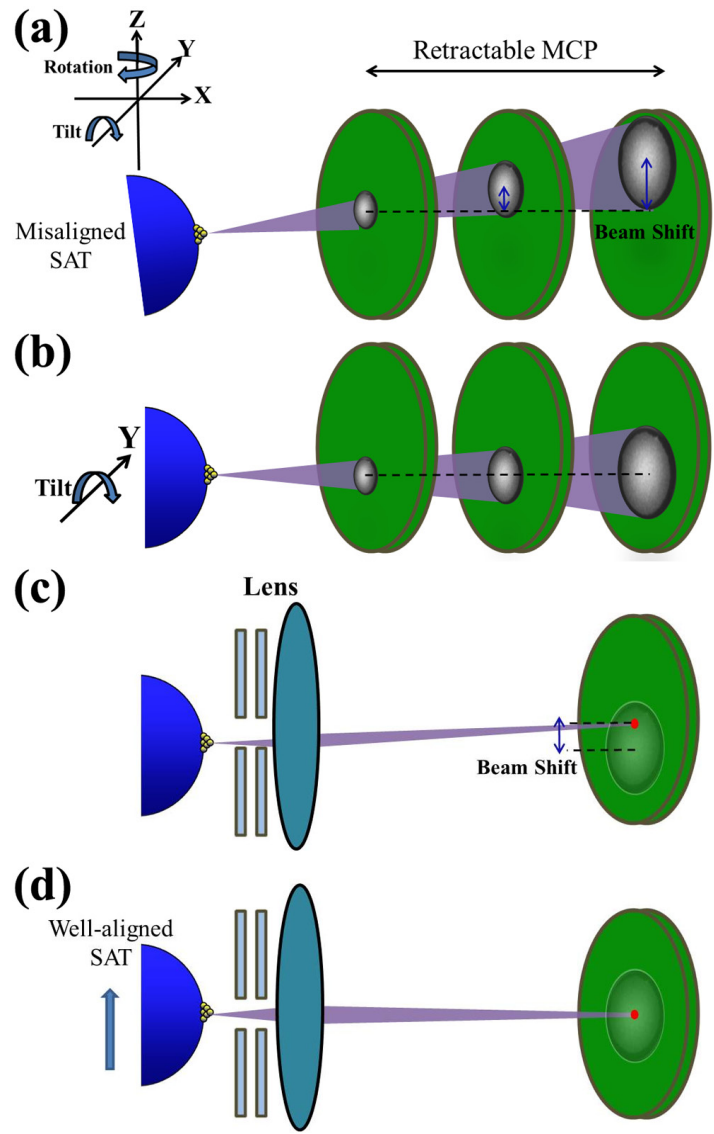

Figure 9  Alignment procedure of a single-atom electron source relative to the lens system. Both the long axis of the lens system and the normal direction of the MCP-screen assembly are along the X axis. (a) Angular misalignment of the electron beam can be inspected by monitoring the beam profile on the MCP at various positions. (b) Misalignment can be compensated using a 2D angular positioner. (c) Lateral misalignment can be inspected by actuating the objective lens. (d) Misalignment is compensated by the multi-axis piezo-positioner.

## 6. Reflection-type coherent diffractive imaging for surface characterization

The mentioned approaches are transmission-type coherent imaging of thin objects. Reflection-type electron diffraction techniques have been widely used to characterize surface structures of numerous materials. In conventional Reflection-type diffraction methods, such as low-energy electron diffraction (LEED) and reflection high-energy electron diffraction (RHEED), the lateral coherence width (tens of nm) of the electron beam is considerably smaller than the illumination width (mm) on the sample. If a fully coherent electron beam can be focused to a sufficiently small size (<100 nm) on a surface, the reflected diffraction patterns can be used in principle to determine the local atomic structures of the surface. This can be a powerful new technique to complement scanning probe microscopy (SPM), such as STM and atomic force microscopy (AFM). STM and AFM have enabled realizing the atomic resolution of surface atomic structures, but only the topmost atomic layer can be imaged. An advantage of coherent electron imaging is that the topmost several layers can be probed with electron waves. Thus reflection-type coherent electron imaging has the potential to be a powerful technique for resolving the three-dimensional atomic structures at the surface.

Figure 10(a) illustrates a schematic of reflection-type CDI. A coherent electron beam is extracted, accelerated, and focused on an area of a sample surface. The reflected diffraction patterns are recorded for reconstructing the surface structures. The electron energy can be tuned to image surface structures of various thickness. Another schematic is illustrated in Figure 10(b), which resembles the configuration of conventional LEED to record backscattered diffraction patterns. To achieve a favorable signal-to-noise ratio in the backscattered diffraction patterns, the electron energy should be less than 200 eV. Achieving favorable focusing with a typical lens is difficult. Researchers have used a microlens to extract and focus electron beams emitted from nanotips (Steinwand et al. 2010; Kratschmer et al. 1995). This can be a solution to achieve favorable focusing at such low electron energies. Single-exposure CDI can be performed to image a local area or ptychographic CDI can be implemented to image an extended surface area. The techniques can also be applied to imaging individual molecules on a crystalline substrate. To improve the diffraction signals, retarded-field grids must be placed in front of the 2D detector to remove the contribution from inelastic scattering.

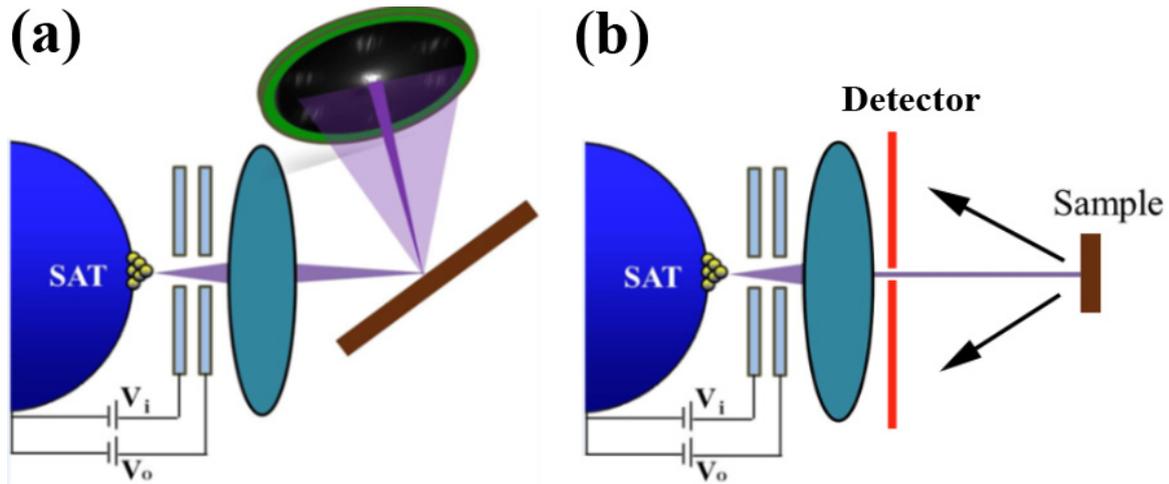

Figure 10 Reflection-type coherent electron imaging for characterizing surface structures. (a) Schematic of a reflection-type CDI. (b) Another schematic to record the backscattered diffraction patterns.

One can also apply a single-atom electron source to low energy electron microscopy (LEEM), which has become a powerful analytical surface science technique (Bauer, 1994; Tromp, 2000). LEEM provides video-rate imaging with 5 nm resolution, which allows observation of dynamic process at surface, such as thin-film grow, strain relief, etching and adsorption, and phase transitions. Graphene growth on specific substrates has also been investigated with this technique (Sutter et al. 2009). If a fully coherent electron source is integrated with LEEM, this may open up a possibility of sub-nanometer or atomic resolution imaging of the sample surfaces by using CDI.

## 7. Conclusion

We propose several experimental schemes to implement low-energy coherent electron imaging on the basis of a highly coherent single-atom electron source. We demonstrate a lens-less transmission-type scheme with operation voltage of 20-500 V. The design is simple and the microscope exhibits high contrast on monolayer graphene. However, the method can be applied only to materials less than 1 nm thick. Furthermore, we propose a transmission-type electron microscope with an operation voltage of 1-10 kV for in-line holography, CDI, and SEM. The electron detector is a retractable image plate used to record the transmission patterns at different positions along the beam direction and to facilitate the process of aligning the electron source relative to the lens system. We also propose reflection-type coherent imaging schemes for characterizing surface atomic structures or the atomic structure of nanostructures or organic molecules on crystalline solid surfaces. Theoretical reconstruction methods must be developed for these schemes. The atomic structures of thin materials, biological molecules, clusters, and material surfaces may be determined, which would be of great scientific importance.

**Acknowledgements**

This research is supported by Academia Sinica of R. O. C. (AS-99-TP-A02).

# References

Abbey B., Nugent K.A., Williams G.J., Clark J.N., Peele A.G., Pfeifer M.A., Jonge M.D., Mcnulty I. (2008) Keyhole coherent diffractive imaging. Nat. Phys. **4**, 394-398.

Algara-Siller G., Lehtinen O., Turchanin A., Kaiser U. (2014) Dry-cleaning of graphene, Appl. Phys. Lett. **104**, 153115

Bauer E. (1994) Low energy electron microscopy. Rep. Prog. Phys. **57**, 895-938.

Bell D.C., Erdman N. (2013) Low voltage electron microscopy principles and applications, first ed., John Wiley & Sons, New Jersey.

Beyer A., Gölzhäuser A. (2010) Low energy electron point source microscopy: beyond imaging. J. Phys.: Condens. Matter **22**, 343001

Binh V.T., Semet V., Garcia N. (1994) Low-energy-electron diffraction by nano-objecs in projection microscopy without magnetic shielding. Appl. Phys. Lett. **65**, 2493.

Binh V.T., Vincent P., Feschet F. (2000) Local analysis of the morphological properties of single-wall carbon nanotubes by Fresnel projection microscop., J. Appl. Phys. **88**, 3385.

Caro L.D., Carlino E., Caputo G., Cozzoli P.D., Giannin C. (2010) Electron diffractive imaging of oxygen atoms in nanocrystals at sub- sub-ångström resolution. Nat. Nanotechnol. **5**, 360-365.

Chang C.C., Kuo H.S., Hwang I.S., Tsong T.T. (2009) A fully coherent electron beam from a noble-metal covered W(111) single-atom emitter. Nanotechnology **20**, 115401.

Chang W.T., Hwang I.S., Chang M.T., Lin C.Y., Hsu W.H., Hou J.L. (2012) Method of electrochemical etching of tungsten tips with controllable profiles. Rev. Sci. Instrum **83**, 083704.

Dellby N., Bacon N.J., Hrncirik P., Murfitt M.F., Skone G.S., Szilagyi Z.S., Krivanek O.L. (2011) Dedicated STEM for 200 to 40 keV operation. Eur. Phys. J. Appl. Phys. **54**, 33505.

Drummy L.F., Yang J., Martin D.C. (2004) Low-voltage electron microscopy of polymer and organic molecular thin films. Ultramicroscopy **99**, 247-256.

Eisele A., Völkel B., Grunze M., Gölzhäuser A. (2008) Nanometer resolution holography with the low energy electron point source microscope. Z. Phys. Chem. **222**, 779-787.

Faulkner H.M.L., Rodenburg J.M. (2004) Movable aperture lensless transmission microscopy: a novel phase retrieval algorithm. Phys. Rev. Lett. **93**,023903.

Faulkner H.M.L., Rodenburg J.M. (2005) Error tolerance of an iterative phase retrieval algorithm for moveable illumination microscopy. Ultramicroscopy **103**, 153-164.

Fienup J.R. (1982) Phase retrieval algorithms: a comparison. Appl Optics **21**, 2758-2769.

Fink H.W., Schönenberger C. (1999) Electrical conduction through DNA molecules Nature **398**, 407-410.

Fink H.W., Stocker W., Schmid H. (1990 a) Coherent point source electron beam. J. Vac. Sci. Technol. B **8**, 1323.

Fink H.W., Stocker W., Schmid H. (1990 b) Holography with low-energy electrons. Phys. Rev. Lett. **65**, 1204.

Fu T.Y., Cheng L.C., Nien C.H., Tsong T.T. (2001) Method of creating a Pd-covered single-atom sharp W pyramidal tip: mechanism and energetics of its formation. Phys. Rev. B **64**, 113401.

Gabor D. (1948) A new microscopic principle, Nature **161**, 777-778.

Germann M., Latychevskaia T., Escher C., Fink H.W. (2010) Nondestructive imaging of individual biomolecules. Phys. Rev. Lett. **104**, 095501.

Glaeser R.M., Taylor K.A. (1978) Radiation damage relative to transmission electron microscopy of biological specimens at low temperature: a review. Journal of microscopy **112,** 127-138.

Huang L.W., Chang C.K., Chien F.C., Chen K.H., Chen P., Chen F.R., Chang C.S. (2014) Characterization of the cleaning process on a transferred graphene. J. Vac. Sci. Technol. A **32**, 050601.

Huang W.J., Zuo J.M., Jiang B., Kwon K.W., Shim M. (2009) sub-ångström-resolution diffractive imaging of single nanocrystals. Nat. Phys. **5**, 129-133.

Humphry M.J., Kraus B., Hurst A.C., Maiden A.M., Rodenburg J.M. (2012) Ptychographic electron microscopy using high-angle dark-field scattering for sub-nanometre resolution imaging, Nature commu. **3**, 730.

Hwang I.S., Chang C.C., Lu C.H., Liu S.C., Chang Y.C., Lee T.K., Jeng H.T., Kuo H.S., Lin C.Y., Chang C.S., Tsong T.T. (2013) Investigation of single-walled carbon nanotubes with a low-energy electron point projection microscope. New J. Phys. **15**, 043015.

Hwang I.S., Kuo H.S., Chang C.C., Tsong T.T. (2010) Noble-metal coverd W(111) single-atom electron sources, J. Electrochem. Soc. **157,** 7-12.

Kaiser U., Biskupek J., Meyer J.C., Leschner J., Lechner L., Rose H.H., Stoger-Pollach M., Khlobystov A.N., Hartel P., Müller H., Haider M., Eyhusen S., Benner G. (2011)

Transmission electron microscopy at 20 kV for imaging and spectroscopy. Ultramicroscopy **111**, 1239-1246.

Kaiser U., Meyer J., Biskupek J., Leschner J., Khlobystov A.N., Müller H., Hartel P., Haider M., Eyhusen S., Benner G. (2010) High resolution 20kV transmission electron microscopy of nanosystems – first results towards sub Ångstrom low voltage EM (SALVE-Microscopy). Microsc Microanal **16**, 1702-1703.

Kamimura O., Dobashi T., Kawahara K., Abe T., Gohar K. (2010) 10-kV diffractive imaging using newly developed electron diffraction microscope. Ultramicroscopy **110**, 130-133.

Kamimura O., Kawahara K., Doi T., Dobashi T., Abe T., Gohara K. (2008) Diffractive microscopy using 20 kV electron beam for multiwall carbon nanotubes. Appl. Phys. Lett. **92**, 024106.

Kamimura O., Maehara Y., Dobashi T., Kobayashi K., Kitaura R., Shinohara H., Shioya H., Gohara K. (2011) Low voltage electron diffractive imaging of atomic structure in single-wall carbon nanotubes. Appl. Phys. Lett. **98**, 174103

Kiesel H., Renz A., Hasselbach F. (2002) Observation of Hanbury Brown-Twiss anticorrelations for free electrons, Nature **418**, 392394.

Kratschmer E., Kim H.S., Thomson M.G.R., Lee K.Y., Rishton S.A, Yu M.L. and Chang T.H.P. (1995) An electron-beam microcolumn with improved resolution, beam current, and stability. J. Vac. Sci. B **13** 2498.

Krivanek O.L., Dellby N., Murfitt M.F., Chisholm M.F., Pennycook T.J., Suenaga K., Nicolosi V. (2010) Gentle STEM: ADF imaging and EELS at low primary energies. Ultramicrscopy **110**, 935-945.

Kuo H.S., Hwang I.S., Fu T.Y., Lin Y.C., Chang C.C., Tsong T.T. (2006) Noble metal/W(111) single-atom tips and their field electron and ion emission characteristics. Jpn. J. Appl. Phys. **45**, 8972-8983.

Kuo H.S., Hwang I.S., Fu T.Y., Wu J.Y., Chang C.C., Tsong T.T. (2004) Preparation and characterization of single-atom tips. Nano Lett. **4**, 2379-2382.

Latychevskaia T., Fink H.W. (2009) Simultaneous reconstruction of phase and amplitude contrast from a single holographic record. Optics express **17**, 10697-10705.

Latychevskaia T., Longchamp J.N., Fink H.W. (2012) When holography meets coherent diffraction imaging, Optics Express **20**, 28871-28892.

Liche H., Lehmann M. (2008) Electron holography – basics and applications, Rep. Prog. Phys. **71**, 016102.

Lin W.H., Chen T.H., Chang J.K., Taur J.I., Lo Y.Y., Lee W.L., Chang C.S., Su W.B., Wu C.I. (2014) A direct and polymer-free method for transferring graphene grown by chemical vapor deposition to any substrate. ASC Nano **8**, 1784-1791.

Lin Y.C., Lu C.C., Yeh C.H., Jin C., Suenaga K., Chiu P.W. (2012) Graphene annealing: how clean can it be?. Nano Lett. **12**, 414-419.

Longchamp J.N., Escher C., Fink H.N. (2013 a) Ultraclean freestanding graphene by platinum-metal catalysis. J. Vac. Sci. Technol. B **31**, 020605

Longchamp J.N., Latychevskaia T., Escher C., Fink H.W. (2012) Low-energy electron transmission imaging of clusters on free-standing graphene. Appl. Phys. Lett. **101,** 113117.

Longchamp J.N., Latychevskaia T., Escher C., Fink H.W. (2013 b) Graphene unit cell imaging by holographic coherent diffraction. Phys. Rev. Lett. **110**, 255501

Marathe S., Kim S.S., Kim S.N., Kim C., Kang H.C., Nickles P.V., Noh D.Y. (2010) Coherent diffractive surface imaging in reflection geometry, Opt. Exp. **18**, 7253-7262.

Meyer J.C., Eder F., Kurasch S., Skakalova V., Kotakoski J., Park H.J., Roth S., Chuvilin A., Eyhusen S., Benner G., Krasheninnikov A.V., Kaiser U. (2012) Accurate measurement of electron beam induced displacement cross sections for single-layer garphene. Phys. Rev. Lett. **108**, 196102.

Miao J., Charalambous P., Kriz J., Sayre D. (1999) Extending the methodology of X-ray crystallography to allow imaging of micrometer-sized non-crystalline specimens, Nature **400**, 342-344.

Miao J., Kriz J., Sayre D. (2000) The oversampling phasing method. ActaCryst. **56**, 1312-1315.

Miao J., Sandberg R.L., Song C. (2012) Coherent X-ray diffraction imaging, IEEE Sel. Top. Quant. **18**, 399-410.

Miao J., Sayre D., Chapman H.N. (1998) Phase retrieval from the magnitude of the Fourier transforms of nonperiodic objects. J. Opt. Soc. Am. A **15**, 1662-1669.

Morishita S., Yamasaki J., Nakamura K., Kato T., Tanaka N. (2008) Diffractive imaging of the dumbbell structure in silicon by spherical-aberration corrected electron diffraction. Appl. Phys. Lett. **93**, 183103.

Morton G.A., Ramberg E.G. (1939) Point projection electron microscope. Phys. Rev. **56**, 705.

Mutus J.Y., Livadaru L., Robinson J.T., Urban R., Salomons M.H., Cloutier M., Wolkow R.A. (2011) Low-energy electron point projection microscopy of suspended graphen, the ultimate microscope slide. New J. Phys. **13,** 063011.

Park J.Y., Kim S.H., Suh Y.D., Park W.D., Kuk Y. (1999) Low-energy electron point source microscope with position-sensitive electron energy analyzer. Rev. Sci. Instrum. **70**, 4304.

Rodenburg J.M. (2008) Ptychography and related diffractive imaging methods, Adv. Imag. Elect. Phys. **150**, 87-184.

Rodenburg J.M., Faulkner H.M.L. (2004) A phase retrieval algorithm for shifting illumination. Appl. Phys. Lett. **85**, 4795.

Rose H.H. (2009) Future trends in aberration-corrected electron microscopy, Phil. Trans. R. Soc. A **367,** 3809-3823.

Sasaki T., Sawada H, Hosokawa F., Kohno Y., Tomita T., Kaneyama T., Kondo Y., Kimoto K., Sato Y., Suenaga K. (2010) Performance of low-voltage STEM/TEM with delta corrector and cold filed emission gun. J. Electr. Microsc. **59**, s7-s13.

Sasaki T., Sawada H., Hosokawa F., Sato Y., Suenaga K. (2014) Aberration-corrected STEM/TEM imaging at 15 kV. Ultramicroscopy **145,** 50-55.

Sasaki T., Sawada H., Hosokawa F., Shimizu Y., Nakamichi T., Yuasa S., Kawazoe M., Kaneyama T., Kondo Y., Kimot K., Suenaga K. (2011) Performance and application of chromatic/spherical aberration-corrected 30kV transmission electron microscope. Microsc Microanal **17**, 1530-1531.

Scherzer O. (1970) Die Strahlenschädigung der Objekte als Grenze für die hochauflösende Elektronenmikroskopie. Berichte der Bunsengesellschaft für physikalische Chemie **74**, 1154-1167.

Spence J.C.H., Qian W., Zhang X. (1994) Contrast and radiation damage in point-projection electron imaging of purple membrane at 100 V. Ultramicroscopy **55**, 19-23.

Spence J.C.H., Weierstall U., Howells M. (2004) Coherence and sampling requirements for diffractive imaging. Ultramicroscopy **101**, 149-152.

Steinwand E., Longchamp J.N., Fink H.W. (2010) Fabrication and characterization of low aberration micrometer-sized electron lenses. Ultramicroscopy **110**, 1148-1153

Suenaga K., Iizumi Y., Okazaki T. (2011) Single atom spectroscopy with reduced delocalization effect using a 30 kV-STEM. Eur. Phys. J. Appl. Phys. **54,** 33508.

Sutter P., Sadowski J.T., Sutter E. (2009) Graphene on Pt(111): growth and substrate interaction, Phys. Rev. B **80**, 245411.

Tromp R.M. (2000) Low-energy electron microscopy. IBM J. Res. Develop. **44**, 503-516.

Wang D.Y., Huang I., Ho P.H., Li S.S., Yeh Y.C., Wang D.W., Chen W.L., Lee Y.Y., Chang Y.M., Chen C.C., Liang C.T., Chen C.W. (2013) Clean-lifting transfer of large-area residual graphene films. Adv. Mater. **25**, 4521-4526.

Weierstall U., Chen Q., Spence J.C.H., Howells M.R., Isaacson M., Panepucci R.R. (2002) Image reconstruction from electron and X-ray diffraction patterns using iterative algorithms: experiment and simulation. Ultramicroscopy **90**, 171-195.

Weierstall U., Spence J.C.H., Stevensand M., Downing K.H. (1999) Point projection electron imaging of tobacco mosaic virus at 40V electron energy. Micron **30,** 335.

Williams D.B., Carter C.B. (1996) Transmission electron microscopy: basic diffraction imaging spectrometry, Plenum Press, New York.

Williams G.J., Quiney H.M., Dhal B.B., Tran C.Q., Nugent K.A., Peele AG., Paterson D., de Jonge M.D. (2006) Fresnel coherent diffractive imaging. Phys. Rev. Lett. **97**, 025506.

Zangwill Z.A. (1998) Physics at Surface, Cambridge University Press, Cambridge.

Zaniewski A.M., Schriver M., Lee J.G., Crommie M.F., Zettl A. (2013) Electronic and optical properties of metal-nanoparticle filled graphene sandwiches. Appl. Phys. Lett. **102**, 023108.

Zuo J.M., Vartanyants I., Gao M., Zhang R., Nagahara L.A. (2003) Atomic resolution imaging of a carbon nanotube from diffraction intensities. Science **300**, 1419-1421.